\magnification=\magstep1
\font\bigbfont=cmbx10 scaled\magstep1
\font\bigifont=cmti10 scaled\magstep1
\font\bigrfont=cmr10 scaled\magstep1
\vsize = 23.5 truecm
\hsize = 15.5 truecm
\hoffset = .2truein
\baselineskip = 14 truept
\overfullrule = 0pt
\parskip = 3 truept
\def\frac#1#2{{#1\over#2}}

\nopagenumbers
\topinsert
\vskip 3.2 truecm
\endinsert
\centerline{\bigbfont TRANSPORT IN A MESOSCOPIC RING WITH A QUANTUM}
\vskip 6 truept
\centerline{\bigbfont DOT: FROM THE COULOMB BLOCKADE REGIME}
\vskip 6 truept
\centerline{\bigbfont  TO THE KONDO EFFECT}
\vskip 20 truept
\centerline{\bigifont Valeria Ferrari, Guillermo Chiappe}
\vskip 8 truept
\centerline{\bigrfont Departamento de F\'\i sica, Facultad de Ciencias
Exactas y Naturales,}
\vskip 2 truept
\centerline{\bigrfont Universidad de Buenos Aires, Pab.1, Ciudad
Universitaria}
\vskip 2 truept
\centerline{\bigrfont (1428) Ciudad de Buenos Aires, Argentina}
\vskip 14 truept
\centerline{\bigifont Enrique V. Anda}
\vskip 8 truept
\centerline{\bigrfont Departamento de Fisica, Pontificia Universidade
Catolica}
\vskip 2 truept
\centerline{\bigrfont do Rio de Janeiro, Rua Marqu\^es de S\~ao Vicente
225}
\vskip 2 truept
\centerline{\bigrfont CEP 22453-900, Rio de Janeiro, Brazil}
\vskip 1.8 truecm

\centerline{\bf 1.  INTRODUCTION}
\vskip 12 truept

During the last years, the technology has achieved to miniaturize
electronic devices to nanometer scales. In order to understand
the physics of these devices, it was developed a new field of research:
the mesoscopic physics [1]. There is a double  motivation to study these
systems: On one hand, it is
the suspicion that could be possible to
construct new devices that give place to a new electronic based
on the principles of mesoscopic physics [2]. On the other hand, the
electrons in such structures are very
confined  and the electron-electron interaction is very
important [3]. For this reason, these devices have been excellent
tools to probe correlation effects in strongly interacting systems.

It has been observed experimentally that a conductor of mesoscopic
dimensions acts as a small pool where  electrons are confined and where
they enter one by one [4]. These conductors have received the name of
Quantum Dots (QDs) [5]. A QD is an artificial device which typically
consists
of a small
region in a semiconductor material  with a size of approximately
100 nm. Due to its small size, the
allowed energies for the electrons are quantized, forming a
discrete spectrum of quantum states. In a typical transport
experiment with QDs, it is observed a peak
in the current every time one electron is added to the QD [6].

The QD is one of the  central objects of this work, the other one is
the mesoscopic ring. These rings are constructed with metallic or
semiconductor
materials and they can be made of sizes of the order of 2500-5000 \AA. If
one of these rings is threaded by a constant magnetic flux,
a current appears in the ring. This current has a periodic
behavior as a function of the applied flux and  exists in the ground
state of the system [7]. Due to this property, this current has received
the name of persistent current.

The system studied in this work consists of a QD inserted in a
mesoscopic ring threaded by a magnetic flux [8,9]. Our aim is to present a
complete description for this device and to predict the physics of a
experiment with these features.

The present article is organized as follows. In section 2 we present the
model and the methodology we have developed to solve it. By means of that
methodology, we have found patterns of behaviour due both to
electron-electron interaction and to the confinement of the QD as we
explain in section 3. Finally, in section 4 we present our conclusions.

\vskip 28 truept

\centerline{\bf 2.  MODEL}
\vskip 12 truept

To solve the ring with the QD, we separate it into two
parts: one of them is a
small cluster of atoms that contains the QD  (see Figure 1) and the
other one is a chain without interactions with a larger number of
sites. Both systems are solved separately and then, they are  connected to
form
the whole system of the QD inserted in the ring (see Figure 2). In first
place, we obtain the Green's Function (GF) for both systems
separately. In second place, we find the GF of the ring with
the QD ($\hat G$), by solving the Dyson's equation $
\hat G = \hat g + \hat g \hat T \hat G$, where $\hat g$ is the GF of one
of the systems and $\hat T$ is the hopping operator that connects
both systems [10,11]. This operator  depends on the flux, because the
magnetic flux is incorporated as a phase factor in the boundary condition
of the ring.

\topinsert
\input psfig.sty
\centerline{\hskip5mm\psfig{figure=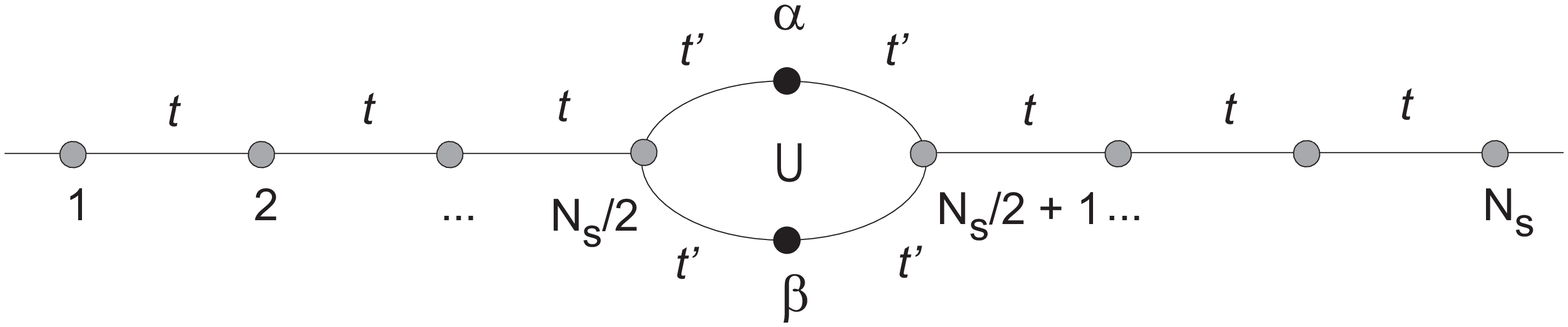,height=11truecm,width=13.5
truecm,angle=0}}
\vskip -3.5truecm
\noindent
{\bf Figure 1.}
Scheme of the $N_s$-site cluster of the atoms that contains the QD.
\vskip 12truept
\endinsert

\topinsert
\input psfig.sty
\centerline{\hskip10mm\psfig{figure=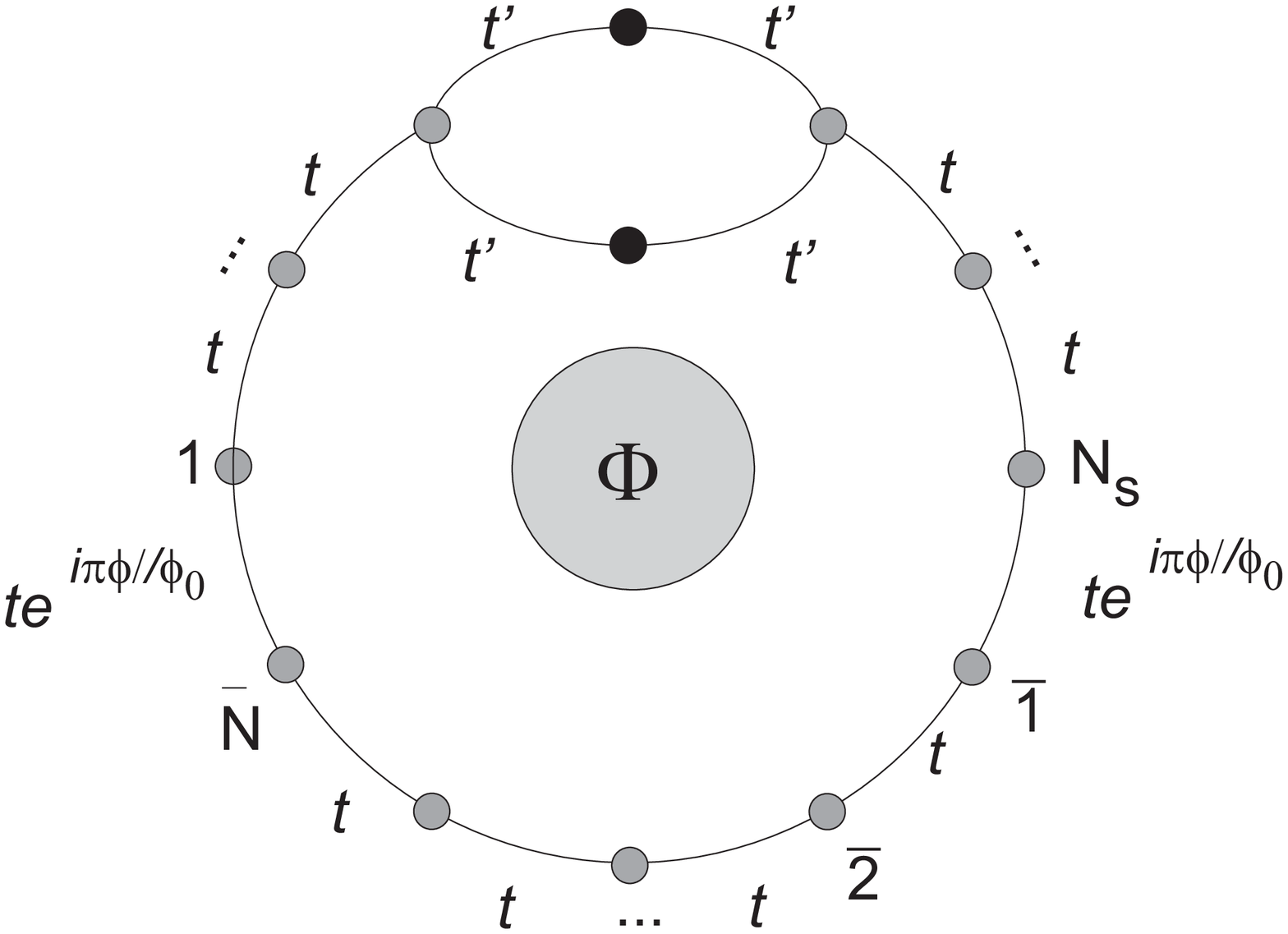,height=12truecm,width=8.5
truecm,angle=0}}
\vskip -2truecm
\noindent
{\bf Figure 2.} Scheme of the mesoscopic ring threaded by a magnetic
flux with an embedded QD. This system is constructed by coupling
the $N_s$-site cluster  to a non-interacting chain of $N$ sites.
\vskip 12truept
\endinsert

We will consider that the QD has two levels: $\alpha$ and $\beta$,
which are connected to the rest
of the system by means of hopping parameter $t'$ (see Figure 1). To model
the QD, this hopping probability for the electrons to hop
to the QD, will be lower than the hopping probability to hop to other
sites of the system ($t´$). Although the Coulomb interaction is present in
the whole system, we will assume it to be restricted to the QD where, due
to quantum confinement, electrons interact more strongly.

In order to reproduce some experiments, we will apply a gate
voltage $V_o$ to the QD. By means of this voltage, we will be
able to move the levels of the QD and study the behavior of
the system as a function of $V_o$.

With this considerations in mind, the Hamiltonian of the cluster
(with $N_s$ sites)  can be written as  sum of three terms: $\hat
H = \hat H_o + \hat H_U + \hat H_T$. $\hat H_o$ represents
essentially the kinetic energy:
$$
\hat H_o = -t \sum_{\scriptstyle \sigma,i = 1 \atop \scriptstyle
i \neq \frac{N_s}{2},\alpha, \beta} ^{N_s}\left[ \hat
c_{i,\sigma}^{\dagger} \hat c_{i+1,\sigma}  + \hat
c_{i+1,\sigma}^{\dagger} \hat c_{i,\sigma} \right] + V_o
\sum_{\scriptstyle \sigma \atop \scriptstyle j=\alpha, \beta}
\hat n_{j,\sigma} + \sum_{\sigma, j=\alpha, \beta} \epsilon_{j}
\hat n_{j,\sigma} $$ \noindent where  $\hat
c_{i,\sigma}^{\dagger}$ ($\hat c_{i,\sigma}$) are the usual
creation (annihilation) fermionic operators ($\sigma \equiv
\uparrow, \downarrow$), $t$ is the hopping parameter between
successive sites,  $\epsilon_{j}$ are the energies of the levels
of the QD (in this work we will take  $\epsilon_{\alpha}=0$ and
$\epsilon_{\beta}=2t$), and $\hat n_{j,\sigma}$ is the number
operator.

The Coulomb interaction inside  the QD is described by
$$\hat H_U = U \sum_{i=\alpha,\beta}  \hat n_{i,\uparrow} \hat
n_{i,\downarrow} + U  \sum_{\sigma=\uparrow, \downarrow} \biggl[ \hat n_{\alpha,\sigma} \hat n_{\beta,-\sigma}  + \hat n_{\alpha,\sigma}
\hat n_{\beta,\sigma}  \biggr]$$
\noindent where  $U$ represents the coulomb energy whenever there are
two electrons inside the QD.

Finally, the connection between the QD and other sites in the cluster is
described by
$$\hat H_T = -t'  \sum_{\sigma=\uparrow, \downarrow} \left[  \hat
c_{\alpha,\sigma}^{\dagger} \hat c_{\frac{N_s}{2}+1,\sigma} +  \hat c_{\alpha,\sigma}^{\dagger} \hat c_{\frac{N_s}{2},\sigma} +\hat
c_{\beta,\sigma}^{\dagger} \hat c_{\frac{N_s}{2}+1,\sigma} +  \hat c_{\beta,\sigma}^{\dagger} \hat c_{
\frac{N_s}{2},\sigma} \right] + c.c. $$
\noindent where $t'$ is the probability for the electrons to hop to the
QD.

To take the charge fluctuations inside the cluster into account, we write
the GF of it, $\hat g$, as a combination of the GF of $n$ and $(n+1)$
particles with corresponding
weights $(1-f)$ and $f$, that is $\hat g= \hat g_n (1-f) + \hat g_{n+1}
f$. The charge inside the cluster is written as $Q_c=(1-f)n+f(n+1)$, but
it can also be expressed as $Q_c=\int_{-\infty}^{\epsilon_F}
\sum_{i}^{N_s}
{\rm Im}G_{ii}(w) {\rm d}w$. These last equations are solved
self-consistently in order to obtain $f$,$n$ and the GF of the system
$\hat G$. Once we get it, we can easily compute the density of
states (DOS), the
persistent current in the ring and the charge inside the QD [10,11].

\vskip 28 truept

\centerline{\bf 3.  PATTERNS OF BEHAVIOUR IN TRANSPORT}
\vskip 12 truept

We have found several patterns of behavior which are evidences
that different physical phenomena are involved in the system. We
have found that we can go from one regime to another either by
diminishing the interaction or by increasing the hybridization
between the QD and the ring. In one extreme, we have found the
Coulomb  Blockade regime and in the other one, we have found a
quasi non interacting regime. Between these two limits,
interesting intermediate regimes appear where the physics is
essentially  dominated by the Kondo effect. In the following, we
will explain in detail every pattern. \vskip 28 truept

\centerline{\bf 3.a Coulomb Blockade regime}
\vskip 12 truept

In this regime, the persistent current and the charge inside
the QD as a function of the gate voltage, looks like in Figure 3.a and
Figure 4.a, respectively. We
can observe a peak in the current every time an electron goes into the
QD and  that there is a range of values of the gate voltage where we have
no transport.

\topinsert
\input psfig.sty
\centerline{\hskip10mm\psfig{figure=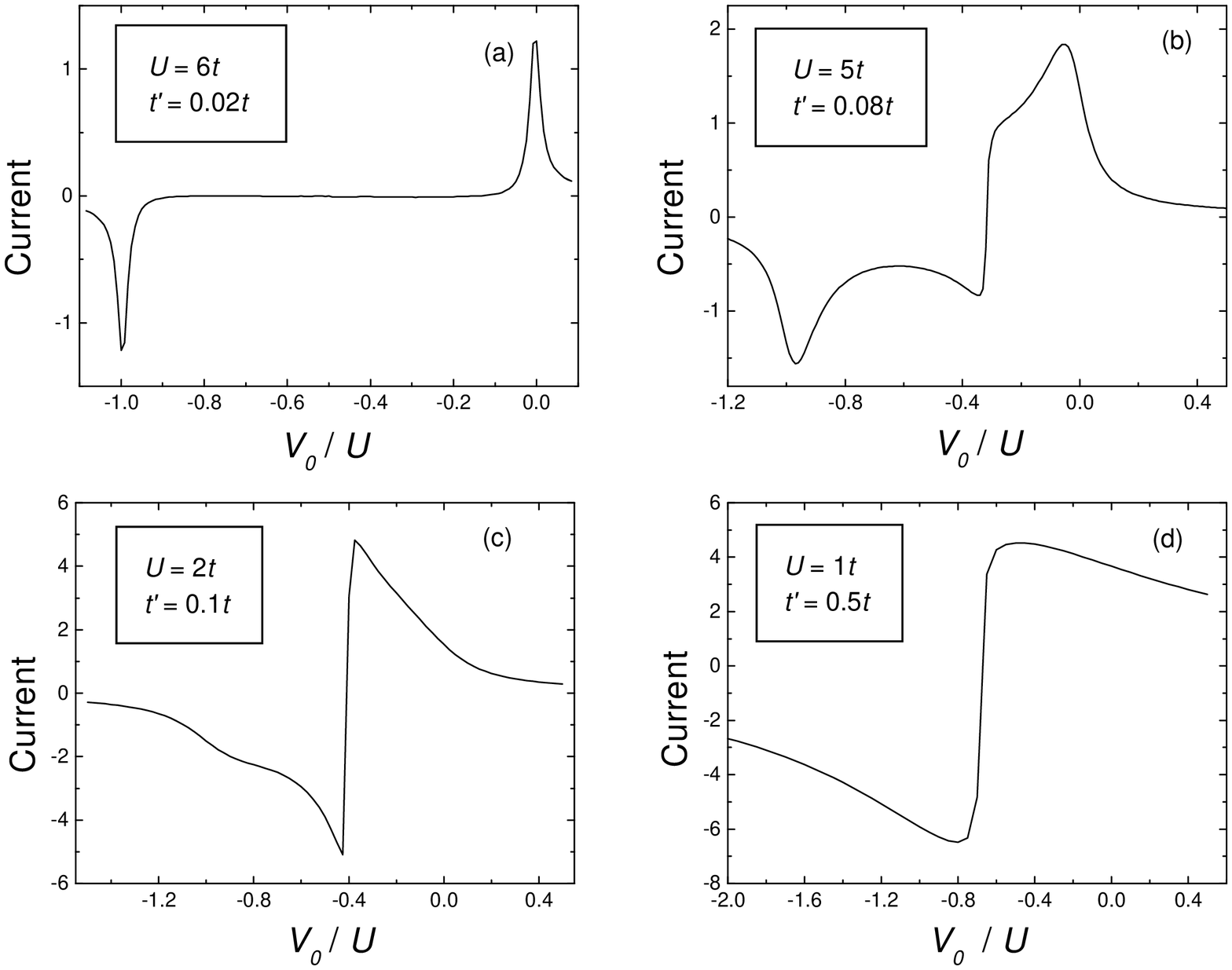,height=17truecm,width=18
truecm,angle=0}}
\vskip -1.5truecm
\noindent {\bf Figure 3.} Persistent current in the ring as a
function of the gate voltage applied to the QD for the different
patterns of behaviour. $(a)$ Coulomb blockade regime. $(b)$
Coulomb blockade-Kondo effect regime. $(c)$ Kondo effect regime.
$(d)$ Quasi non-interacting regime. The gate voltage is normalized
to $V_o/U$ to better compare among different graphs.

\vskip 12truept
\endinsert

\topinsert
\input psfig.sty
\centerline{\hskip10mm\psfig{figure=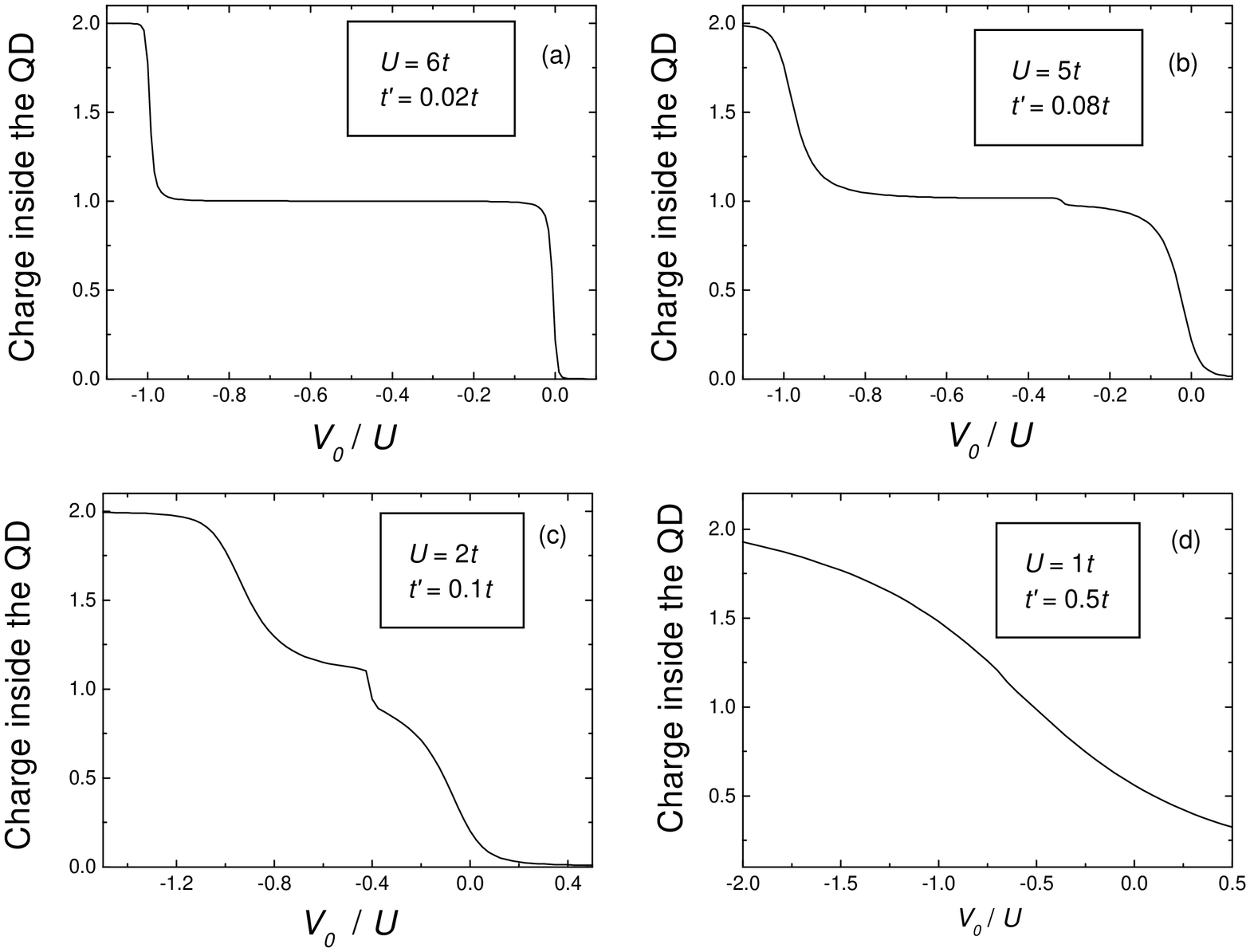,height=17truecm,width=18
truecm,angle=0}}
\vskip -1.5truecm
\noindent
{\bf Figure 4.}
Charge inside the dot as a function of the gate voltage (normalized to
$V_o/U$) for the different patterns of behaviour $(a)$ Coulomb blockade
regime. $(b)$ Coulomb blockade-Kondo effect regime. $(c)$ Kondo effect
regime. $(d)$ Quasi non-interacting regime.
\vskip 12truept
\endinsert

In order to understand this result, it is useful to take into account the
local DOS at the QD. If we consider a value of gate
potential such that there is one particle inside the QD in the level
$\epsilon_{\alpha}$, we will have
two peaks in the DOS: one at  $\epsilon_{\alpha}$
and the other one at $\epsilon_{\alpha}+U$ (indicating that to add another
particle it is necessary an amount of energy  $U$). These peaks in the
DOS have a finite width $\Delta$ and we have  found
that those peaks will broaden out if the hybridization
between the QD and the ring (tuned by $t'$) is increased [11].

With this considerations in mind, we can understand  the graphs
of the persistent current and the charge. Suppose we begin the process
with a value of gate voltage in such a way that
the lowest level of the QD is aligned with the Fermi level of the system.
In this situation the transport is allowed, so we have current
and one electron entering  into the QD, as actually occurs
in  Figure 3.a and Figure 4.a at zero gate voltage.

If we lower the gate voltage, we will move the levels downwards,
so there will be no states of the QD in resonance with the Fermi
level in the ring. If this occurs, the transport is
blocked, and we will have no current. This phenomena is known as
Coulomb blockade [12]. This situation will persist during an interval of
gate voltage of the order of $U$, and after that, the second level
appears and the second particle goes into the QD.

With regard to the parameters of the model, we have found this regime
when correlations are much more greater than the width of the
levels ($\Delta$) due to hybridization.
\vskip 28 truept

\centerline{\bf 3.b Kondo effect - Coulomb blockade regime}
\vskip 12 truept

By diminishing the value of the interaction $U$ - maintaining it
greater than $\Delta$- we find a pattern where Kondo effect and
Coulomb blockade coexist [10].

In this regime, we can note in Figure 3.b  an excess of current
between the  Coulomb Blockade peaks that indicate the entrance of
particles inside the QD. This behaviour can be easily understood
if we note that in the DOS, a peak at the Fermi level appears,
whenever we have an odd number of electrons inside the QD [10].
This peak is  evidence of the Kondo effect which appears in metals
with magnetic impurities and it occurs because the spin of the
magnetic impurity is coupled anti-ferromagneticaly with the spin
of the conduction electrons. It shows up as a resonance in the
Fermi level, which is usually called Kondo resonance (KR) [13].
On the other side, when  there is an even number of electrons,
the resonance at the Fermi level disappears.

When we have an odd number of electrons, there is a net spin
inside the QD, so it behaves as a magnetic impurity whose spin
can be coupled anti-ferromagneticaly with the spins of the
electrons in the ring. In this situation, the QD behaves as a
magnetic impurity and we have Kondo effect. The Kondo resonance at
the Fermi level gives a new channel for the transport and it is
responsible of the excess of current between the CB peaks. On the
other side, when we have an even number of electrons, we have no
net spin inside the QD, so it is no possible to have Kondo
effect. The fact that a QD could exhibit Kondo effect has been
predicted theoretically [14] and it has been observed in recent
experiments [15]. \vskip 28 truept

\centerline{\bf 3.c Kondo effect regime}
\vskip 12 truept

If we diminish  $U$ or increase  $t'$ again, we get the pure
Kondo effect regime. A typical graph for the persistent current in this
regime can be seen in Figure 3.c.

It is interesting to point out that if we increase  $t'$ or diminish
$U$ in such a way that the parameter $J = \frac{4 t'^{2}}{U}$
remains constant, we observe  the same behaviour for the
persistent current in the ring.  This feature can be observed in Figure 5.
The parameter $J$ is essentially the coupling between
the spin of the QD and the electrons in the ring.
In other words, we have found that the parameter $J$ governs
the whole physics in this regime, and it occurs because
in the present regime we have the pure Kondo effect.

\topinsert
\input psfig.sty
\centerline{\hskip15mm\psfig{figure=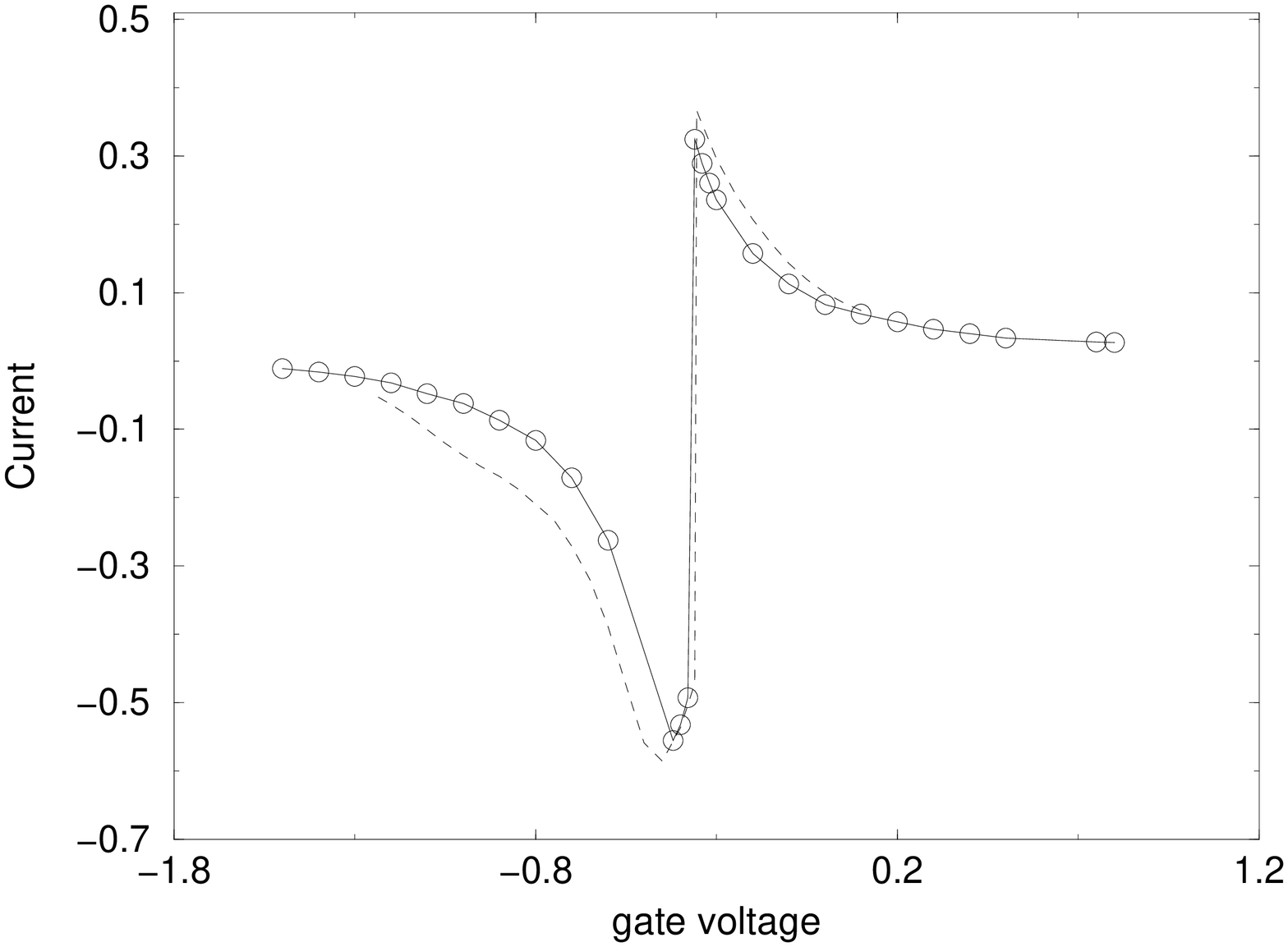,height=10truecm,width=13.5truecm,angle=270}}
\vskip 1truecm
\noindent {\bf Figure 5.} Persistent current as a function of the
gate voltage in the pure Kondo effect regime. The ring is
constructed connecting a 6-sites cluster with a chain of 2000
sites. The dash line is the current for $U=2t$ and $t'=0.16t$ and
the circles correspond to $U=0.5t$ and $t'=0.08t$. In both cases
the value of the coupling parameter is the same, that is:
$J=0.0512t$. \vskip 12truept
\endinsert

In a recent work [10], we have found that the regime where the Kondo
effect coexists with the Coulomb blockade,
the KR is very narrow and as a consequence,  the persistent current
presents an scaling with the length of the ring ($L$) as
$\frac{1}{\sqrt{L}}$. In
the present regime the KR widens, so the current
presents the usual $\frac{1}{L}$ scaling of a perfect ring. However,
as the Kondo temperature can be roughly estimated as the width of the KR
[13], we can conclude that this temperature will be greater than that of
the previous regimes.

In Figure 4.c we can observe  that the
charge begins to lose its discreteness, because charge enters
into the QD in a almost continuous way but  it is  still
entering one by one.

Regarding the parameters of the model, this regime appears when the
coupling $J$ is lower than the width of the peaks in the local DOS [11].
That is, $J = \frac{4{t'}^2}{U} < \Delta$.
\vskip 28 truept

\centerline{\bf 3.d Quasi non-interacting regime}
\vskip 12 truept

For lower values of $U$, we find the quasi
non-interacting regime with a typical curve for the persistent current as
in Figure 3.d.

If $U << \Delta$, the peaks in  the DOS merge and we get
one doubly degenerated peak.
In this regime, the KR is absent because the
charge inside the QD changes in a continuous way
from 0 to 2 particles (as can be seen in Figure 4.d). So, we
can say that the system behaves essentially as if it were
non-interacting.

\vskip 28 truept

\centerline{\bf 4.  CONCLUSIONS}
\vskip 12 truept

In the present work we have studied the transport properties of QD
inserted in a mesoscopic ring which is threaded
 by a magnetic flux.

We have proposed a model that takes into account the  conditions that
are usual in experiments with QDs [8,9]. In order to solve
 the model, we develop a methodology to find the Green's functions
of the system by means of proper approximations.

This system presents persistent currents as a function of a gate
voltage applied to the QD. We have studied  the influence of both
the interaction between electrons and the hybridization between
the ring and the QD. We have found several regimes that describe
different physical phenomena involved in the system. These
regimes range from the phenomenon of Coulomb Blockade (in the
high correlation limit) to a quasi non-interacting regime.
Between these two limits we have found an intermediate regime
where the Kondo effect shows up. Similar results has been
recently reported in other systems [16].

\vskip 28 truept

\centerline{\bf ACKNOWLEDGMENTS}
\vskip 12 truept

V. F. thanks  Eduardo Mucciolo for important suggestions and
acknowledges support from University of Buenos Aires, Argentina.
\vskip 28 truept

\centerline{\bf REFERENCES}
\vskip 12 truept

\item{[1]} Supriyo Datta, {\it Electronic Transport in Mesoscopic
Systems} (Cambridge University Press, 1995); Y. Imry, {\it Introduction to
Mesoscopic Physics} (Oxford University Press, 1997); T. Ando, Y. Arakawa,
K. Furuya, S. Komiyama, and H. Nakashima,  {\it Mesoscopic Physics and
Electronics} (Springer-Verlag, Berlin, 1998).

\item{[2]} T. J. Thornton, Rep. Prog. Phys. {\bf 57}, 311-364 (1994).

\item{[3]} P. Lee in {\it Transport Phenomena in Mesoscopic Systems},
Springer Series in Solid State Sciences
109. Edited by H. Fukuyama and T. Ando (Berlin, Springer, 1992) page 31.

\item{[4]}  M.Kastner, Rev. Mod. Phys. {\bf 64}, 849 (1992); M.
Devoret, D. Esteve, and C. Urbina, Nature (London) {\bf 360}, 547 (1992);
M. Devoret and H. Grabert in {\it Single Charge Tunneling:
Coulomb Blockade Phenomena in Nanostructures}, NATO ASI Series B 294.
Edited by H. Grabert, M. Devoret (Plenum, New York, 1992); M. Devoret, D.
Esteve, and C. Urbina, in {\it Single electron phenomena in metallic
nanostructures}. Edited by E. Akkermans, G. Montambaux, J. Pichard, and
J. Zinn-Justin (Elsevier, Amsterdam, 1995).

\item{[5]} M. Kastner, Physics Today {\bf 46}, 24 (1993); P. McEuen,
Science {\bf 278}, 1682 (1997); L. Kouwenhoven, Ch. Marcus, P. McEuen, S.
Tarucha, R. Westervelt, and N. Wingreen, {\it Electron Transport in
Quantum Dots}, Proceedings of the Advanced Study Institute on {\it
Mesoscopic Electron  Transport}. Edited by L. Sohn, L. Kouwenhoven and
G. Sch\"on. (Kluwer, 1997); L. Kouwenhoven and Ch. Marcus, Physics World,
35, June (1998).

\item{[6]} R. C. Ashoori, Nature (London) $\bf 379$, 413 (1996).

\item{[7]} L. L\'evy, G. Dolan, J. Dunsmuir, and H. Bouchiat, Phys.
 Rev. Lett. {\bf 64}, 2074 (1990); V. Chandrasekhar, R. Webb, M. Brady, M.
Ketchen, W. Gallagher, and A. Kleinsasser,
Phys. Rev. Lett. {\bf 67}, 3578 (1991); D. Mailly, C. Chapelier, and A.
Benoit, Phys. Rev. Lett. {\bf 70}, 2020 (1993); B. Reulet, M. Ramin, H.
Bouchiat, and D. Mailly, Phys. Rev. Lett. {\bf 75}, 124 (1995).

\item{[8]} A. Yacoby, M. Heiblum, H. Strikman, and D. Mahalu,
Phys. Rev. Lett. {\bf 74}, 4047 (1995).

\item{[9]} W. G. van der Wiel, S. De Franceschi, T. Fujisawa,
J.M.Elzerman, S.Tarucha, and L. P.Kouwenhoven, Science {\bf 289}, 2105
(2000).

\item{[10]} V. Ferrari, G. Chiappe, E. Anda, and Maria
Davidovich, Phys. Rev. Lett. {\bf 82}, 5088 (1999).

\item{[11]} V. Ferrari, Ph.D. Thesis, University of Buenos Aires (1999).

\item{[12]}  H. van Houten and C. Beenakker, Phys. Rev. Lett. {\bf 63},
1893 (1989); C. Beenakker, Phys. Rev. {\bf B 44}, 1646 (1991).

\item{[13]}  A. C. Hewson, {\it The Kondo Problem to Heavy Fermions},
(Cambridge University Press, 1993).

\item{[14]} T. Ng and P. Lee, Phys. Rev. Lett. {\bf 61}, 1768 (1988); L.
I. Glazman and M. Raikh,  JETP Lett. {\bf 47}, 452 (1988); Y. Meir,
N. Wingreen, and P. Lee, Phys. Rev. lett. {\bf 66} 3048 (1991); T.
Inoshita, A. Shimizu, Y.Kuramoto, and H. Sakaki, Phys. Rev. {\bf B 48},
14725 (1993); Y. Meir, N. Wingreen, and  P.Lee, Phys. Rev. lett. {\bf 70}
2601 (1993);  N. Wingreen and Y. Meir, Phys. Rev. {\bf B 49}, 11040
(1994); Y. Wan, P. Phillips, and Q. Li, Phys. Rev. {\bf 51} R14782 (1995).

\item{[15]} D. Goldhaber-Gordon, Hadas Shtrikmna, D. Mahalu,
D. Abusch-Magder, and M. Kastner,  Nature (London) {\bf 391}, 156 (1998);
S. M. Cronenwett, T. Ooterkamp, L. Kouwenhoven, Science {\bf 281}, 540
(1998); S. Sasaki, S. De Franceschi, J.M. Elzerman, W.G. van der Wiel, M.
Eto, S. Tarucha, and L.P. Kouwenhoven, Nature (London) {\bf 405}, 764
(2000).

\item{[16]} A. Levy Yeyati, F. Flores, and A. Mart\'\i n-Rodero, Phys.
Rev. Lett. {\bf 83}, 600 (1999); Shi-Jie Xiong and Ye Xiong, Phys. Rev.
Lett. {\bf 83}, 1407 (1999).

\end